\documentclass[12pt,a4paper,onecolumn]{IEEEtran} 
\usepackage{euscript}
\usepackage{cite}
\usepackage{amsmath,amssymb,amsbsy,mathrsfs,flushend}
\usepackage{mathtools,bbm,bm}
\usepackage{color,pifont}
\usepackage{graphicx}
\usepackage{booktabs} 
\usepackage{float} 
\usepackage{tikz,circuitikz}
\usepackage{caption,subcaption}
\usepackage{theorem}
\usepackage{fancyhdr}
\usepackage{url}
\usepackage[pagebackref=true]{hyperref}

\usepackage{hyperref}
\def\beq{\begin{equation}}
\def\eeq{\end{equation}}

\newcommand{\HRule}{\rule{\linewidth}{0.2mm}}
\renewcommand\descriptionlabel[1]%
      {\hspace{\labelsep}\textsf{#1}}
\def\insertfig#1#2#3#4#5{
\begin{figure}[#1]
\centering
{\includegraphics[width=#2\columnwidth,clip=]{#3.pdf}}
\caption{#4}\label{#5}\end{figure}}
\begin{document}
\title{\begin{LARGE}
\HRule \\[0.4cm]
Line codes generated by finite Coxeter groups
\end{LARGE}
      }

\author{Ezio~Biglieri,~\IEEEmembership{Life~Fellow,~IEEE,}
Emanuele~Viterbo,~\IEEEmembership{Fellow,~IEEE}%
\thanks{The work of E.\ B.\ was supported by Project~TEC2015-66228-P, while that of
E.\ V.\  was supported by ARC under Grant Discovery Project No. DP160101077.}%
\thanks{Preliminary versions of this manuscript were presented in~\cite{bigvit1,bigvit2}.}
\thanks{Ezio Biglieri is with the Electrical Engineering Department, UCLA, and with Departament TIC, Universitat Pompeu Fabra, Barcelona, Spain (email: {\tt e.biglieri@ieee.org}). Emanuele Viterbo is with the Department of Electrical and Computer Systems Engineering, Monash University,
Melbourne, Australia (email: {\tt emanuele.viterbo@monash.edu}).}
\vspace{-1cm}
}
\maketitle
\noindent
\HRule \\[0cm]

\begin{abstract}
\noindent
Using an algebraic approach based on the theory of Coxeter groups, we design, and describe the performance of, a class of line codes for parallel transmission of $b$ bits over $b+1$ wires that admit especially simple encoding and decoding algorithms. A number of designs are exhibited, some of them being novel or improving on previously obtained codes.
\end{abstract}
\begin{IEEEkeywords}
Line coding, group codes for the Gaussian channels, permutation modulation, Coxeter groups.
\end{IEEEkeywords}
%
%
\section{Introduction and motivation of the work}
In this paper we describe the design of vector line codes allowing an especially simple maximum-likelihood (ML) detection procedure. This consists of a linear transformation of the vector received at the output of an additive white Gaussian noise (AWGN) channel, followed by a binary slicer. The design is based on the selection of a subset of a permutation modulation (PM) codebook being the direct product of binary antipodal
signaling schemes, and hence having a geometrical representation in the form of a multidimensional {\em orthotope} (or hyper-rectangle). The encoder can also be implemented as a linear transformation of the source (binary) vector.

Transmission on parallel wireline links (as those used to interconnect integrated circuits, or a television set to a set-top box) is affected by disturbances placing a number of constraints on the design of the signaling scheme. The key problem here is the design of line codes allowing the transmission of $b$ bits over $w \geqslant b$ wires and using a codebook $\EuScript W$ subject to some constraints to be detailed later. The general scheme is shown in Fig.~\ref{figure1bis}.
\insertfig{h!}{0.9}{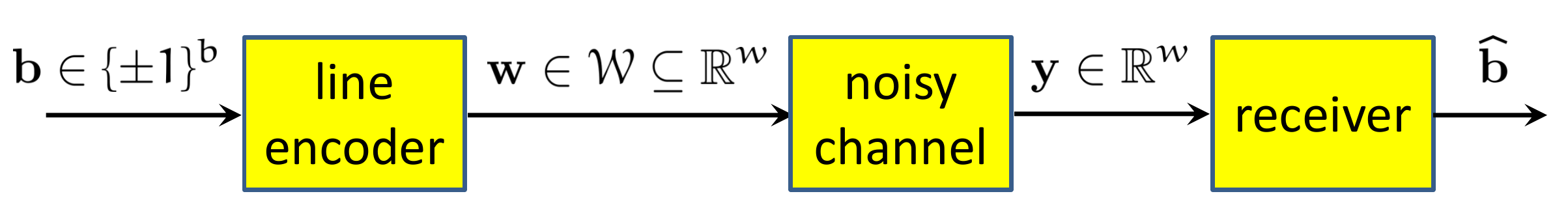}{\sl General scheme of vector coding.}{figure1bis}
Here, $b$ binary information symbols $\pm 1$ are input in parallel to the $(w,b)$ line encoder, which is a one-to-one map from $\{ \pm 1 \}^b$ to the codebook $\EuScript W \subseteq {\mathbb R}^w$. This encoder outputs
a code vector $\bf  w$ with $w$ real components, which is added to white Gaussian noise to
obtain vector $\bf y$. Vector $\bf y$ is processed by a detector whose output is an estimate $\widehat{\bf b}$ of the information vector.

\begin{figure}[h!]
\centering
\begin{subfigure}[b]{0.5\textwidth}
\centering
\begin{circuitikz}[scale=0.7]
\ctikzset{voltage/bump b=25pt,voltage/european label distance=20pt}
\draw
(2,1.22)to[short,i=$\mbox{$i\in \{0,I\}$}$](6,1.22)|- (6,1.22)  to[short,i=$$] (8,2.5)
 (8,-0.5) to[R,v=$\mbox{$v\in \{0,RI\}$}$,l=$R$]  (8,2.5)
 (8,-0.5) node[ground]{}
;
\end{circuitikz}
\caption{\sl Binary unipolar signaling (NRZ): $(1,1)$ line code. }
\end{subfigure}
\vspace{5mm}

 \begin{subfigure}[b]{0.5\textwidth}
 \centering
\begin{circuitikz}[scale=0.7]
\ctikzset{voltage/bump b=25pt,voltage/european label distance=20pt}
\draw
(2,2.1)to[short,i=$\mbox{$i\in \{+I,-I\}$}$](6,2.1)|- (6,2.1)  to[short,i=$$] (8,3.5)
 (8,0) to[R,v=$\mbox{$v\in \{+RI,-RI\}$}$,l=$R$]  (8,3.5)
 (8,0) to[short,i=$$] (6,1.35)
|- (6,1.35)to[short,i=$$](2,1.35)
;
\end{circuitikz}
\caption{\sl Binary differential signaling (DS): $(2,1)$ line code.}
\end{subfigure}
\caption{\sl Binary unipolar (NRZ) and binary
differential signaling (DS) circuits. \label{fig:1_1vs2_1}}
\end{figure}
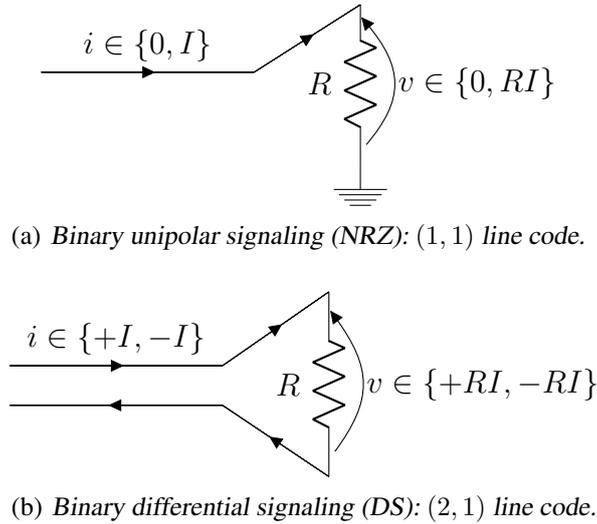

Fig.~\ref{fig:1_1vs2_1} illustrates the two basic circuits for wired binary communications:
(a) unipolar signaling and (b) differential signaling.
The transmitter sends a current signal $i$ through one wire in (a) and two wires in (b), and the receiver measures a voltage $v$ across the resistor $R$. In~(a) a threshold voltage $V_t=RI/2$
is used by a comparator to detect the binary information, while in~(b) a zero-threshold voltage $V_t=0$ is used.
Since the power dissipated on the resistor $R$ is $Ri^2$, on the average (a) uses half of the power of (b), but reduces by a factor of two the distance to the threshold.
Since the thermal noise produced by the resistor is the same for both (a) and (b),
this implies a power gain of 1.5dB of~(b) over~(a) to achieve the same performance.

Other types of impairments may heavily affect the reliability of unipolar signaling:
({\em i})  Coupling of EM interference with the transmission wire (common-mode disturbances), since the return current goes through the ground plane, and
({\em ii}) Power supply fluctuations due to simultaneous switching noise
(SSN), which affect the stability of the threshold voltage.
On the other hand, differential signaling provides common-mode rejection of ({\em i}), and the zero threshold is insensitive to ({\em ii}) because the total current drawn from the power supply is kept constant (for further details see, for example, [1], [7], [9]).
The advantages of DS come at the price of a reduction of the wire efficiency from $1$~bit/wire to $0.5$~bit/wire, and an increase of the complexity needed by the transmitter to drive the currents.
Recent work (partially listed among the References below) has focused on the design of signaling schemes that retain the advantages of DS while improving wire efficiency.

The circuit for a $(3,2)$ line code receiver was illustrated in~\cite{bigvit1}.
In general, the receiver is realized by a star of $w$ resistors with a common center node,
where the transmitted zero-sum currents converge to reduce the overall SSN.
The $w$ codeword components represent current signals at the transmitter on $w$ wires, and the
receiver uses zero-voltage comparators across ${w}\choose{2}$ resistor pairs.
These comparators determine the sign of the differences between all pairs of components of the received vector (as illustrated in next Section) in order to provide the sorting order of these components.

In this paper we take an algebraic\slash geometric approach to the design and analysis of line coding schemes transmitting $b$ bits over $w=b+1$ wires. We show how a matrix with orthogonal rows can transform a suitable subset of a PM
vector set~\cite{slepian} into a signal constellation whose geometric representation in the Euclidean space is a $b$-dimensional orthotope, which leads to a simple ML decoding algorithm based on one binary slicer per wire. Our design accounts for several types of impairments that may be present besides additive white Gaussian noise. Specifically, resistance to common-mode noise is obtained
by using code words whose components are {\em balanced} (i.e., sum to zero), simultaneous switching output noise is reduced by using constant-energy signals, and the effects of intersymbol interference are reduced by having
only two amplitude values at the input of each slicer~\cite{primer,patcronie2}.
Codebook design is based on the theory of Group Codes for the Gaussian Channel~\cite{slepian2},
 as specialized to groups generated by reflections in orthogonal hyperplanes.

After examining a simple example for motivation (Section~\ref{motivation}), a general theory
is expounded in Section~\ref{theory}. Design examples are shown in Section~\ref{examples}, while performance evaluation is presented in Section~\ref{performance}. Section~\ref{optimization} deals with the optimization of the design, and additional remarks are presented in Section~\ref{miscel}.
\section{An example for motivation and illustration}\label{motivation}
An early approach to the code design problem described in previous Section was taken in~\cite{ohetal}, where
a coding scheme based on a number of wires greater than $2$ and having a wire efficiency $2/3$~bit/wire was advocated. This scheme was generalized by Abbasfar in~\cite{abbasfar}, where a multiwire (``vector'') DS scheme was designed. Under the assumption
that the transmitted amplitudes are $\pm 1$, the number of $+1$ (and hence of $-1$) in all transmitted vectors is kept constant, which makes this signaling scheme balanced. An example of this {\em generalized differential vector signaling} scheme is provided by the following set of $6$ vectors (the codebook) used for transmission of $\log_2 6$~bits over $w=4$ wires. Exhibiting the codebook in the form of a matrix whose rows are the codewords, we have
\beq\label{PM1}
{\bf W} = \left[
\begin{array}{rrrr}
 +1&-1&+1&-1\\
-1&+1&+1&-1\\
-1&-1&+1&+1\\
+1&-1&-1&+1\\
-1&+1&-1&+1\\
+1&+1&-1&-1 \end{array}\right]
\eeq
Vectors~\eqref{PM1} form a Variant-I PM set~\cite{ericson,slepian,vitperm}, obtained as the set of all the permutations of an initial vector $(-1,-1,+1,+1)$. In general, a (Variant-I) PM codebook
in ${\mathbb R}^n$ is obtained as the
set of all the distinct permutations of an initial $n$-vector ${\bf w}_1$. Assuming that
${\bf w}_1$ has $r$ distinct components with multiplicities $m_1, \ldots, m_r$, and $\sum_{i=1}^r m_i=n$, these permutations are in number of
$n!/(m_1!m_2!\cdots m_r!)$.
A peculiar feature of PM is that optimum (ML) detection over the additive white Gaussian noise (AWGN) channel is especially simple.
In fact, to decode the received $n$-vector $\bf y$, the ML receiver need only arrange its coordinates in decreasing order. This is equivalent to finding the signs of the ${n}\choose{2}$ differences
between the components of $\bf y$, and comparing these signs with the entries of a lookup table (notice also that the requirement of balanced vectors in the codebook words leads to the optimality of the PM scheme, in the sense discussed in~\cite{bigeli}---more on this in Section~\ref{optimization}).

Now, line codes based on the PM scheme may be improved upon if a codebook $\EuScript W$ can be found such that:
({\em i})~It includes a number of codewords equal to a power of $2$, so that $|{\EuScript W}|=2^b$, ({\em ii})~Only $b$ signs of linear expressions need be computed for ML detection, ({\em iii})~These signs are the source symbols, so that no lookup table is needed by the decoder, and ({\em iv})~Encoding can be obtained by a linear operation on source symbols. Some line codes satisfying conditions ({\em i})--({\em iv}) were designed by Shokrollahi {\em et al.} (see  \cite{abbasfar,primer,patcronie1,patcronie2,patshok1} and references within). In this paper we derive a general theory of these codes, based on the concepts of Group Codes for the Gaussian Channel and of Coxeter groups.

We start with a relatively simple design example whose illustration will motivate the theory developed in the balance of this paper.
Consider the PM codebook with $6$ words\footnote{In the following, we shall use interchangeably the terms word, vector, or point, to describe one element of $\EuScript W$.} obtained as all the permutation of the components of the initial vector ${\bf w}_1=(-1,0,1)$. Denoting by $i$-$j$ the  difference between the $i$th and the $j$th components of a vector, in the absence of noise the word is identified by the signs of the differences $1$-$2$, $2$-$3$, and $1$-$3$ between pairs of its components. The situation is summarized
in Table~\ref{basictable1}, where those differences are shown for all codebook vectors,
as received in the absence of noise.
\begin{table}[H]
\centering
\caption{\sl Vectors of a PM codebook and differences between their components.}
\label{basictable1}
\begin{tabular}{@{}cc|ccc@{}}
\addlinespace[15pt]
\toprule
& vector & $1$-$2$ & $2$-$3$ & $1$-$3$ \\
  \addlinespace[5pt]
\midrule
\addlinespace[5pt]
 \ding{192} & $(-1,0,1)$ & $-1$ & $-1$ & $-2$   \\
 \ding{193} & $(-1,1,0)$ & $-2$ & $+1$ & $-1$   \\
 \ding{194} & $(0,-1,1)$ & $+1$ & $-2$ & $-1$   \\
 \ding{195} & $(1,0,-1)$ & $+1$ & $+1$ & $+2$   \\
 \ding{196} & $(0,1,-1)$ & $-1$ & $+2$ & $+1$   \\
 \ding{197} & $(1,-1,0)$ & $+2$ & $-1$ & $+1$  \\
\bottomrule
\end{tabular}
\end{table}
We can interpret the operations summarized in Table~\ref{basictable1} as a linear mapping $\EuScript L$ between the original codebook $\EuScript W$ and its transformed version
${\EuScript W}^\prime\triangleq{\EuScript L}\EuScript W$, whose words are listed in the right part of Table~\ref{basictable1}. Since this transformation is one-to-one, ${\EuScript W}^\prime$ can be detected in lieu of the original codebook $\EuScript W$. For this observation to be practically useful, we need to consider PM schemes
such that decoding ${\EuScript W}^\prime$ in the presence of noise is equivalent to decoding ${\EuScript W}$,
but simpler. The simplest situation, which is the one on which we shall focus
our attention in the balance of this paper, occurs when the words of codebook ${\EuScript W}^\prime$
(interpreted as points in the $3$-dimensional Euclidean space ${\mathbb R}^3$) are vertices of a $3$-orthotope. If all these vertices are included (which is obtained when the number of codewords chosen is a power of $2$, viz., $2^b$), then  ${\EuScript W}^\prime$ {\em can be optimally detected by simply taking the sign of each entry} of the transformed vector ${\EuScript L}{\bf y}$, i.e., feeding it to a slicer. The slicer outputs are elements of $\{ \pm 1 \}^b$.
Thus, the ML receiver in this situation consists
of a linear transformation $\EuScript L$ followed by $b$ slicers. This fact can also be used for {\em encoding} purposes: in fact, even encoding can be done linearly, by applying a suitable linear transformation to any vector containing, in an appropriate form to be described later, $b$ entries of the form $(\pm 1, \pm 1, \ldots, \pm 1)$.

We now proceed to explain in detail how the concept above can be implemented. The tips of the $6$ vectors of Table~\ref{basictable1} are the vertices of a regular hexagon lying on the surface of a $3$-dimensional sphere with radius $\sqrt{2}$, as shown in Fig.~\ref{esagono_sfera_ter}.
\insertfig{h!}{0.4}{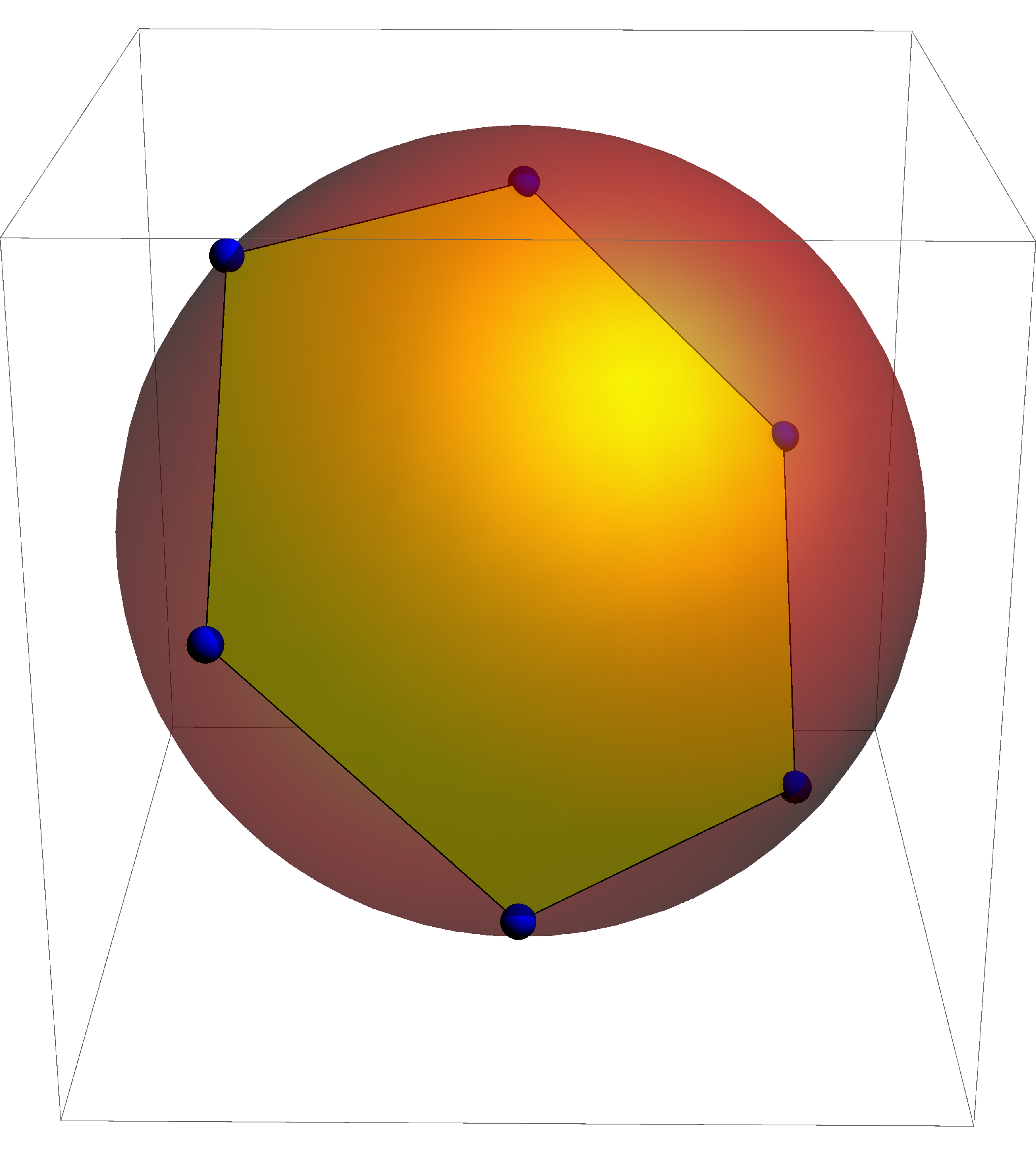}{\sl Geometric representation of codebook in Table~\ref{basictable1}.}{esagono_sfera_ter}
Since all points of ${\bf w} \in \EuScript W$ lie on the hyperplane $\langle {\bf w} , {\bf 1}\rangle=0$,
where $\bf 1$ denotes the vector all of whose components are $1$, we may
project the points of $\EuScript W$ on this plane to obtain a $2$-dimensional representation. A general way of performing this projection was described by Peterson in~\cite{peterson}. The projection of the $n$-vector $\bf w$ on the plane described by the scalar product $\langle {\bf w} , {\bf 1}\rangle=0$ is obtained by computing ${\bf wA}$, where $\bf A$ is the $n\times n$ projection matrix
\beq\label{petmat}
{\bf A} = \left[ \begin{array}{cccccc}
1+\beta &\beta &\beta &\cdots&\beta &\gamma \\
\beta &1+\beta &\beta &\cdots&\beta &\gamma \\
 & & \ddots & & & \\
 \beta &\beta &\beta & \cdots &1+\beta &\gamma \\
 \gamma & \gamma & \gamma &\cdots &\gamma & \gamma
\end{array}\right]
\eeq
where $\gamma \triangleq 1/\sqrt{n}$ and $\beta \triangleq -1/(n-\sqrt{n})$.
Using this with $n=3$, the code vectors are transformed into vectors whose third component is zero, thus reducing the codebook representation to a $2$-dimensional space, as illustrated in Fig.~\ref{esagono2D}.
\insertfig{h!}{0.4}{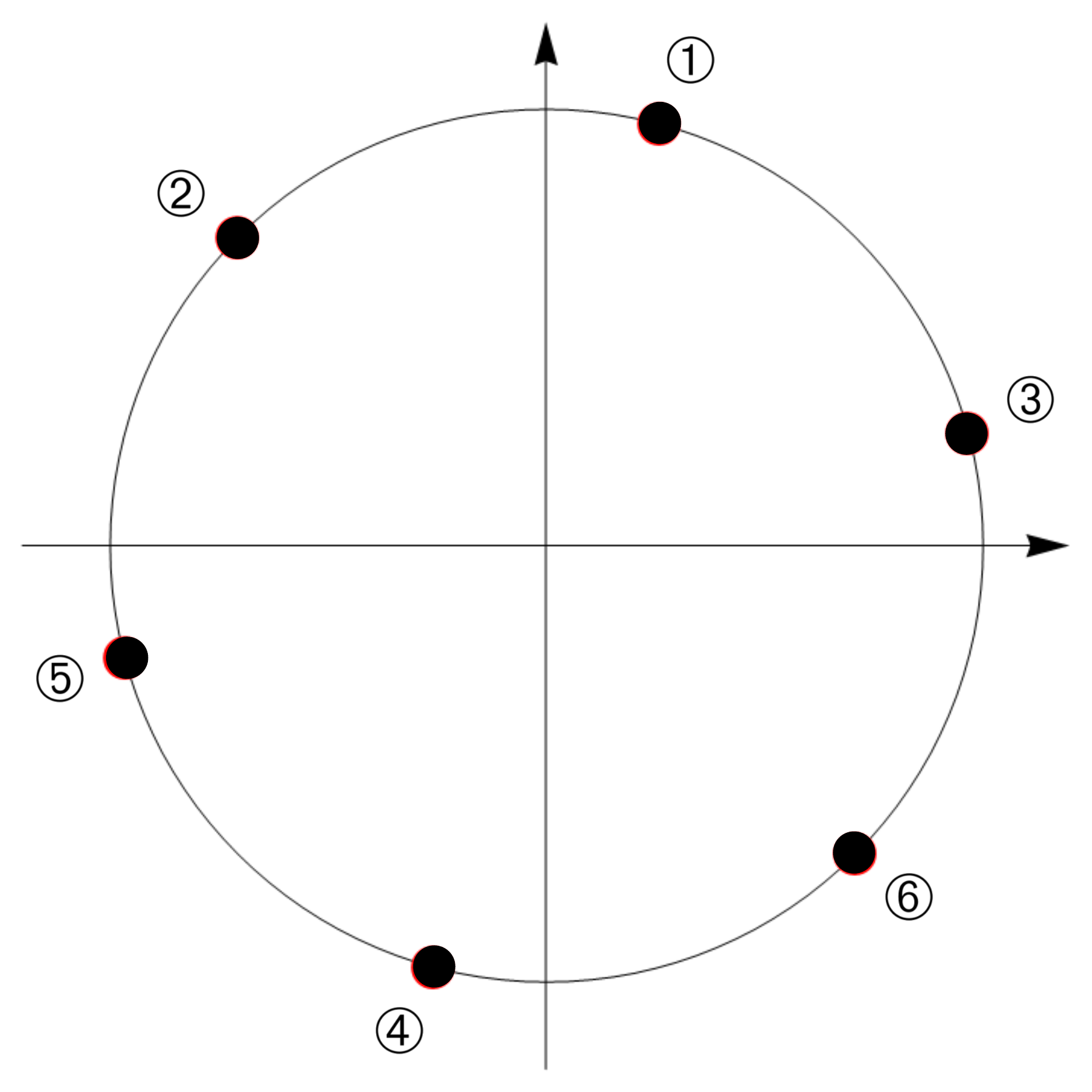}{\sl $2$-dimensional geometric representation of codebook in Table~\ref{basictable1}.}{esagono2D}

In the specific case we are now handling, the reduced codebook with $2^b=4$ words
whose geometric representation has a rectangular shape is
obtained by removing vectors
\ding{194} and \ding{196} from the $6$-vector PM set of Table~\ref{basictable1}. In matrix form:
\beq\label{simpleexample}
{\bf W} =
\left[ \begin{array}{rrr} -1&0&1\\-1&1&0\\1&0&-1\\1&-1&0 \end{array} \right]
\hspace{-5pt} \begin{array}{c} \mbox{\ding{192}} \\ \mbox{\ding{193}} \\ \mbox{\ding{195}} \\  \mbox{\ding{197}}  \end{array}
\eeq
The ML (congruent) decision regions of this codebook are defined by their boundary planes, as obtained from the equations
\beq\label{decreg}
\langle {\bf y} , ({\bf w}_i - {\bf w}_j)\rangle = 0
\eeq
where ${\bf w}_i$, ${\bf w}_j$ are neighbors. Eq.~\eqref{decreg} expresses the fact that the separating plane is orthogonal to the line
joining ${\bf w}_i$ and ${\bf w}_j$, or, equivalently, that $\bf y$ has the same distance from
${\bf w}_i$ and ${\bf w}_j$. The plane separating neighbors ${\bf w}_1$ and ${\bf w}_2$ has equation $y_2-y_3=0$, while that separating ${\bf w}_1$ and ${\bf w}_6$ has equation
$2y_1-y_2-y_3=0$, or, equivalently, $(y_1-y_2)+(y_1-y_3)=0$ (Fig.~\ref{figuronaredrawed}).
\insertfig{h!}{0.4}{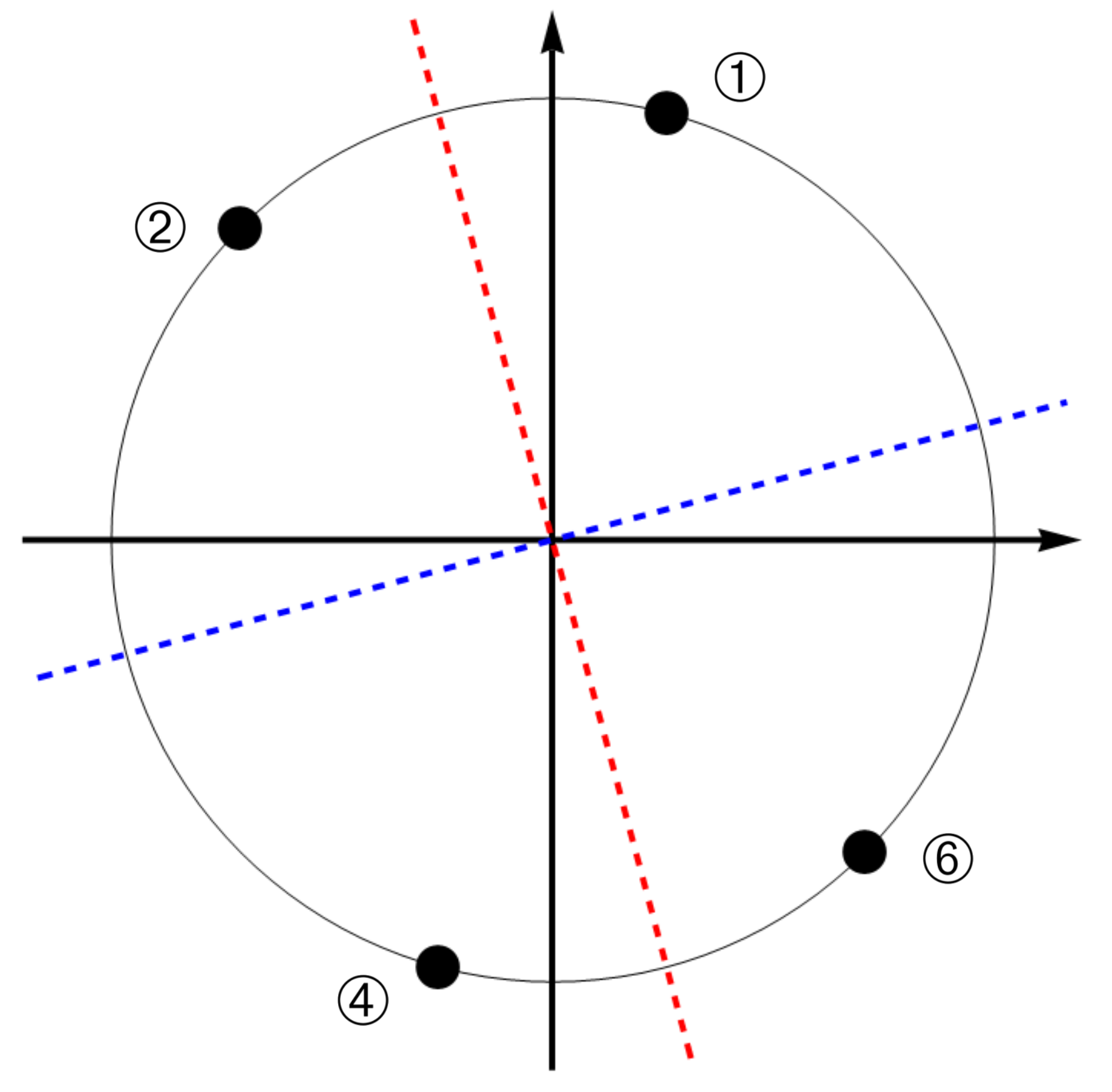}{\sl $2$-dimensional representation of codebook
\eqref{simpleexample}. Dashed lines: Separators of ML decision regions based on the signs of $2$-$3$ and of $(1$-$2)+(1$-$3)$.}{figuronaredrawed}

The ML detection procedure is summarized in Table~\ref{tavolachespiega}.
\begin{table}[H]
\centering
\caption{\sl Transformation of
codebook $\bf W$ into
onto the vector set ${\bf W}^\prime$ with $4$ elements $\pm 3, \pm 1$, the four vertices of a rectangle.}
\label{tavolachespiega}
\centering
\begin{tabular}{@{}ccccccc@{}}
\addlinespace[15pt]
\toprule
& vector & $(1$-$2)+(1$-$3)$ & $2$-$3$  \\
  \addlinespace[5pt]
\toprule
\addlinespace[5pt]
\ding{192} & $(-1,0,1)$ & $-3$ & $-1$     \\
 \ding{193} & $(-1,1,0)$ & $-3$&  $\phantom{-}1$     \\
 \ding{195} & $(1,0,-1)$ & $\phantom{-}3$ &  $\phantom{-}1$    \\
 \ding{197} & $(1,-1,0)$ & $\phantom{-}3$ &   $-1$     \\
\bottomrule
\end{tabular}
\end{table}
This Table describes the transformation $\EuScript L$ which maps the
codebook $\bf W$
onto the vector set ${\bf W}^\prime$ with $4$ elements $\pm 3, \pm 1$, the four vertices of a rectangle. The decision regions of the transformed codebook are delimited by the coordinate axes in the $2$-dimensional plane, and hence the transmitted vector can be detected by simply slicing the components of ${\EuScript L}{\bf y}$, as indicated above. The linear transformation of $\bf W$ is given by the {\em detection matrix}
\beq\label{mat1}
{\bf M} = \left[ \begin{array}{rrr} 1&1&1 \\ 2 &-1 &-1  \\ 0&1&-1  \end{array}
\right]
\eeq
which has orthogonal rows (notice that $\bf M$ itself is not orthogonal, so that transformation by $\bf M$ alters the scales of the coordinate axes).
Its first row reflects the fact that the sum of the components of each row of $\bf W$ is zero (balanced codewords), the second row corresponds to the difference $(1$-$2)+(1$-$3)$, and the third row to the difference $(2$-$3)$. Thus, we have
\beq\label{mat2}
{\bf WM}^{\sf T} = \left[ \begin{array}{rrr}
0&-3 &-1 \\ 0&-3 &1 \\ 0&3 &1 \\ 0&3 &-1  \end{array}
\right]
\eeq
which expresses the transformation of the original vector set $\bf W$ into the new vector set
whose two-dimensional representation is shown in Fig.~\ref{rettangolo}. In turn, this vector set can be detected using simply the signs of its second and third components.
\insertfig{h!}{0.4}{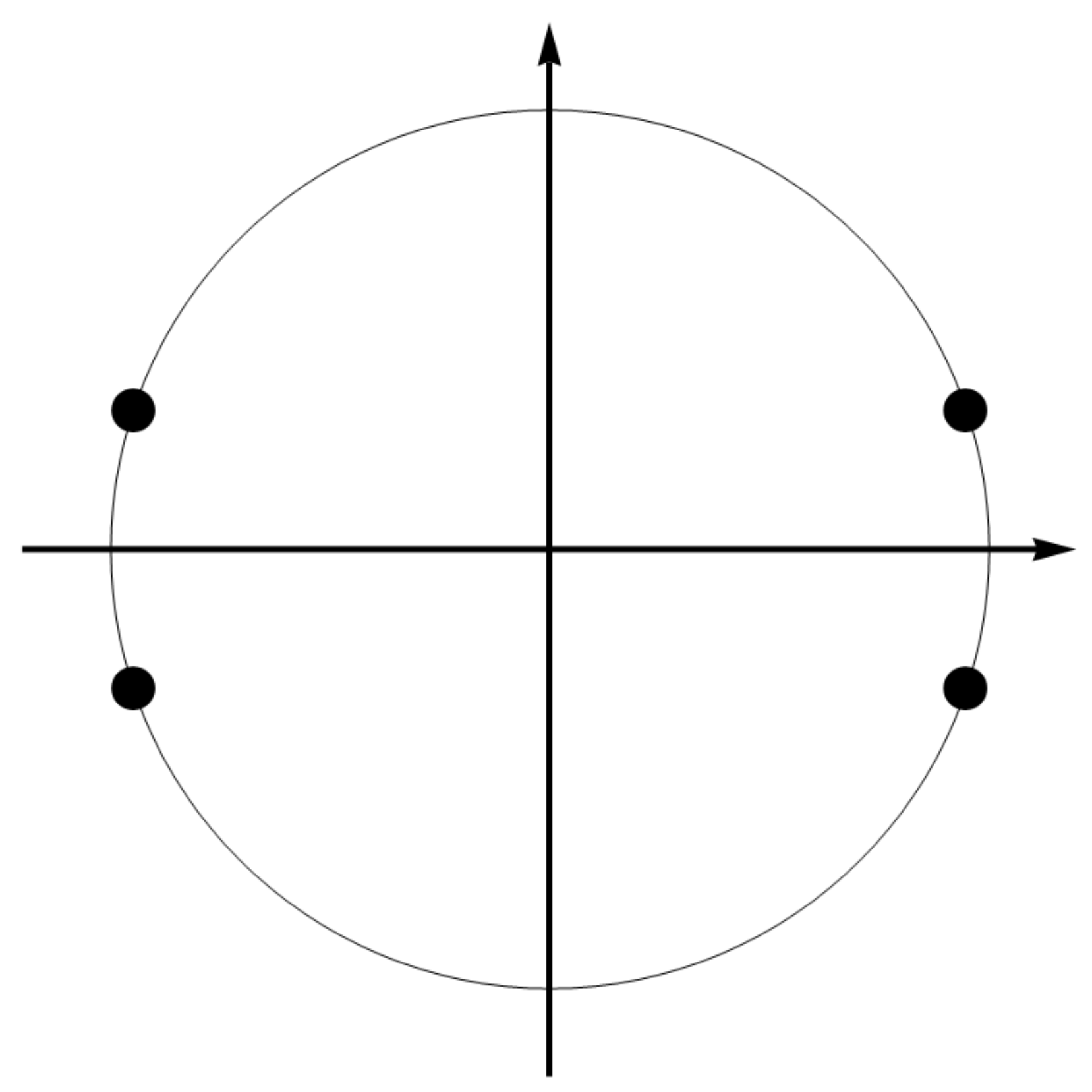}{\sl $2$-dimensional representation of the linearly transformed codebook ${\bf WM}^{\sf T}$.}{rettangolo}
Conversely, the linear coding procedure transforms the {\em information matrix} \beq
{\bf B} = \mbox{sgn}({\bf WM}^{\sf T} ) = \left[ \begin{array}{rrr} 0&-1&-1 \\ 0&-1&1 \\ 0&1&1 \\ 0&1&-1 \end{array}
\right]
\eeq
into $\bf W$ as
\beq
{\bf W} =  {\bf BK}
\eeq
where ${\bf K}= \frac{1}{2} {\bf M}$. Notice that $\EuScript L$ transforms the $3$-dimensional codebook $\bf W$  into the $2$-dimensional codebook ${\bf W}^\prime$. Thus, the inverse transformation mapping the $4$ information symbols
into $\bf W$ should be a map between a $2$-dimensional and a $3$-dimensional space. This explains why each row of $\bf B$ has a ``zero'' prepended.
\subsection{Towards a general theory}
We now describe how codebook~\eqref{simpleexample} can be generated directly, i.e., without necessarily thinking of it as a subset of a PM set. The basic requirement here is that $\EuScript W$ be a {\em group code for the Gaussian channel}~\cite{slepian2}.
This is generated by the action of a group of real orthogonal matrices ${\bf O}_i$, $i=1,\ldots,2^b$, on an initial vector ${\bf w}_1$, and results into Voronoi regions that are congruent. The geometric structure we are interested in is that of an orthotope, which suggests that the Voronoi regions be bounded by orthogonal hyperplanes. In the example we are examining, the two matrices
\beq\label{genmat1}
{\bf O}_1=\left[ \begin{array}{ccc} 1&0&0\\0&0&1\\0&1&0 \end{array}\right]
\qquad
{\bf O}_2= \frac{1}{3} \left[ \begin{array}{rrr} -1&2&2\\2&2&-1\\2&-1&2
\end{array}\right]
\eeq
satisfy the condition ${\bf O}_1^2={\bf O}_2^2= {\bf I}$, and represent reflections
in the orthogonal planes with normal vectors ${\bf w}_1-{\bf w}_2=(0,-1,1)$ and
${\bf w}_1-{\bf w}_6 =(-2,1,1)$, respectively.
They commute, and $({\bf O}_1{\bf O}_2)^2=1$. Thus, ${\bf O}_1$ and ${\bf O}_2$ generate the matrix group of order $4$ with elements ${\bf I},{\bf O}_1,{\bf O}_1{\bf O}_2, {\bf O}_2$. The product of these matrices by the initial vector $(-1,0,1)$ yields the codebook~\eqref{simpleexample}.
\section{Basic theory}\label{theory}
Being guided by the considerations developed in Section~\ref{motivation}, we now expound the general theory leading to line codes in the shape of orthotopes. The appropriate mathematical tool is the theory of finite Coxeter groups. These~\cite{bjobre,coxeter,coxeterbook,groben,humphreys} are defined by generators and relations. A finite Coxeter group ${\EuScript G}$ has a presentation with generator set $\EuScript R = \{ R_i \}$ and relations $(R_iR_j)^{m(i,j)}=I$, where $R_i$, $R_j$ are group elements, $I$ denotes the identity element of $\EuScript G$, and $m(i,j)$ are integers. In particular, $m(i,i)=1$, and $m(i,j)=2$ if and only if $R_i$ and $R_j$ commute. A Coxeter group has a convenient description in terms of a graph having as nodes the elements of $\EuScript R$ and as edges the unordered pairs $\{ R_i, R_j \}$ such that $m(i,j)\ge 3$. The edges with $m(i,j)\ge 4$ are labeled by that number. The group is irreducible if its Coxeter graph is connected. For example, the graph with $n$ isolated nodes
\beq\label{cox}
\raisebox{-.5ex}{$\bullet$} \phantom{\rule{1cm}{0.4pt}} \raisebox{-.5ex}{$\bullet$} \phantom{\rule{1cm}{0.4pt}} \raisebox{-.5ex}{$\bullet$}
\; \; \raisebox{-.5ex}{$\cdots$} \; \; \raisebox{-.5ex}{$\bullet$} \phantom{\rule{1cm}{0.4pt}} \raisebox{-.5ex}{$\bullet$}
\eeq
is the Coxeter graph of a group isomorphic to ${\mathbb Z}_2^n$ of order $2^n$. The graph
\beq
\raisebox{-.5ex}{$\bullet$} \rule{1cm}{0.4pt} \raisebox{-.5ex}{$\bullet$} \rule{1cm}{0.4pt} \raisebox{-.5ex}{$\bullet$}
\; \; \raisebox{-.5ex}{$\cdots$} \; \; \raisebox{-.5ex}{$\bullet$} \rule{1cm}{0.4pt} \raisebox{-.5ex}{$\bullet$}
\eeq
with nodes labeled $R_1$, $R_2$, \dots, $R_{n-1}$, is the Coxeter graph of the symmetric group $S_n$ whose generators are the adjacent transpositions $R_i=(i,i+1)$, $1\le i \le n$.

A {\rm reflection} on the Euclidean space ${\mathbb R}^n$ is a linear transformation
of ${\mathbb R}^n$ of codimension $1$, called its {\em mirror} and having a nontrivial eigenvector with eigenvalue $-1$, called a {\em root} of the reflection. A reflection can be represented by the matrix ${\bf I}-2{\bm \delta} {\bm \delta}^{\sf T}$, where $\bm \delta$ is the corresponding
unit-norm root vector normal to the mirror plane. Coxeter groups are generated by reflections, so that each node in the Coxeter graph corresponds to a reflection. In particular, when $m(i,j)=2$ the corresponding reflections are in orthogonal planes. Thus, we are especially interested in groups of the form~\eqref{cox} whose elements
are faithfully represented by $(b+1)\times (b+1)$ matrices corresponding to reflections in hyperplanes that are mutually orthogonal, as well as orthogonal to the
hyperplane whose normal vector is the all-one vector $\bf 1$. Once such a group with order $2^b$ is found, the line code
$\EuScript W$ is obtained by applying the matrix group to an initial $(b+1)$-vector ${\bf w}_1$. As a result, we obtain a line code whose
$2^b$ Voronoi regions are congruent and bounded by orthogonal hyperplanes. In geometric terms, the line code turns out to be equivalent to a Cartesian product of binary antipodal signals, with an added dimension allowing the codewords to be balanced.

To generate from a PM set the Coxeter group we need, we advocate the following procedure. Start from a balanced initial $(b+1)$-vector ${\bf w}_1$, and choose $b$ {\em root permutations}\footnote{With an abuse of terminology, we identify a permutation with the vector obtained by permuting the components of ${\bf w}_1$.} ${\bf w}_i$, $i=2, \ldots, b+1$, of ${\bf w}_1$ such that
the $b$ unit-norm (column) vectors
\beq\label{getdelta}
{\boldsymbol \delta}_i \triangleq \frac{{\bf w}_1-{\bf w}_{(i+1)}}{
                                  \| {\bf w}_1-{\bf w}_{(i+1)} \| }, \qquad i=1, \ldots, b
\eeq
are mutually orthogonal. (Section~\ref{optimization} describes an algorithm to select these root permutations.)
Next, the corresponding reflection matrices
\beq
{\bf O}_i \triangleq {\bf I} - 2{\boldsymbol \delta}_i{\boldsymbol \delta}_i^{\sf T}
\eeq
are computed.
Direct calculation shows that ${\bf O}_i^2=({\bf O}_i{\bf O}_j)^2= {\bf I}$,
so that these matrices generate a Coxeter matrix group isomorphic to a power of ${\mathbb Z}_2$. The group code $\EuScript W$ is obtained
by applying this matrix group to ${\bf w}_1$~\cite{slepian2}. From now on, we shall describe the group code by listing its vectors as rows of the $2^b\times (b+1)$ matrix $\bf W$.

Since the Voronoi region associated with a point ${\bf w}_i\in {\EuScript W}$ is the set of points lying closer to
${\bf w}_i$ than to any other ${\bf w}_j$, $j\ne i$, from the equality $\| {\bf y} - {\bf w}_i\|=
\| {\bf y} - {\bf w}_j\|$ we see that $\langle {\bf y} , ({\bf w}_i-{\bf w}_j)\rangle=0$ defines the hyperplane
halfway between ${\bf w}_i$ and ${\bf w}_j$  and orthogonal to the vector
$({\bf w}_i-{\bf w}_j)$ (see Fig.~\ref{voronoibis}).
%
\insertfig{h!}{0.50}{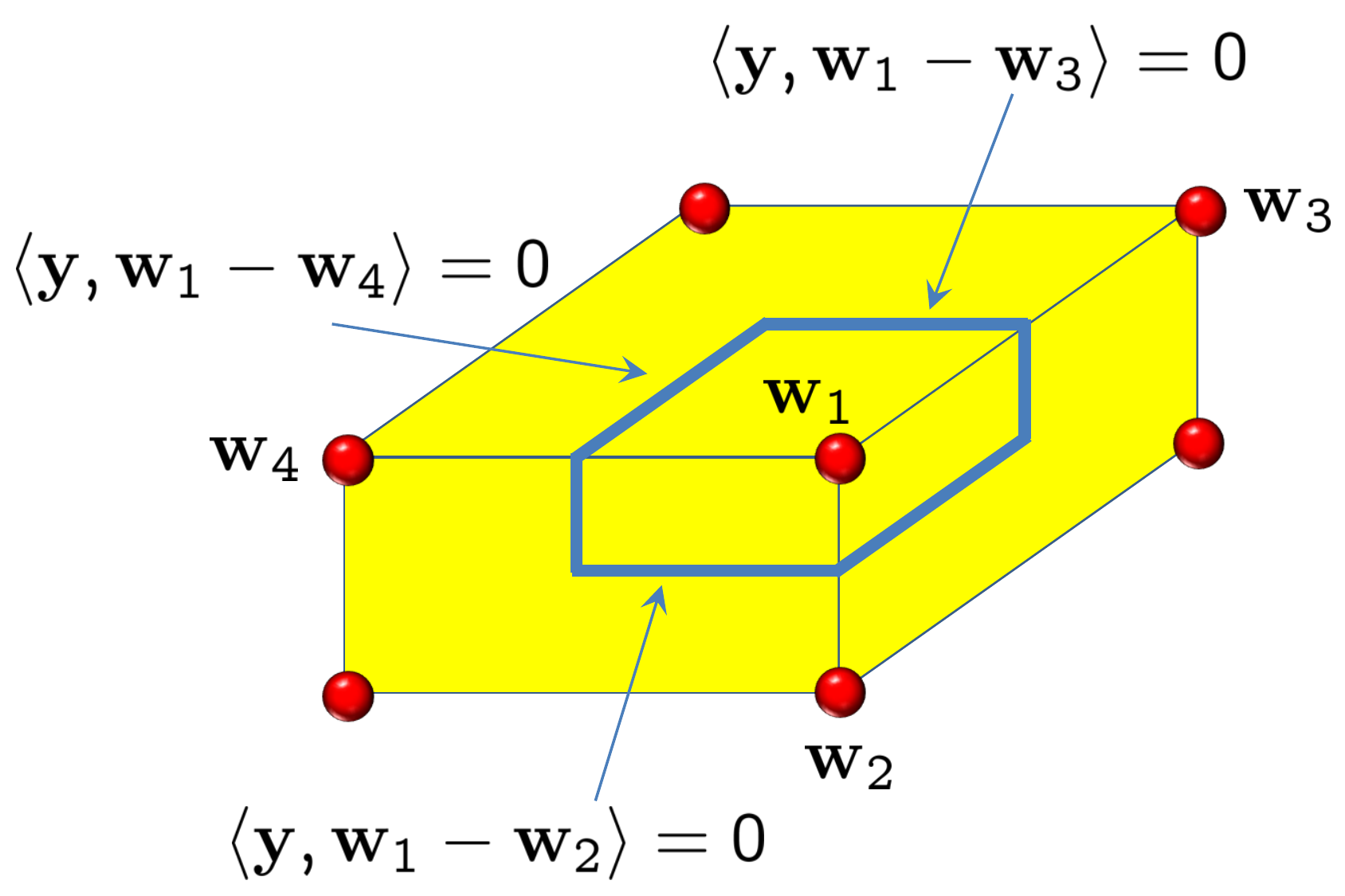}{\sl $3$-dimensional Voronoi region for a orthotope-shaped codebook.}{voronoibis}

The scalar product is positive if $\bf y$ is closer to ${\bf w}_i$ than to ${\bf w}_j$, and vice versa. Thus, defining a matrix $\bf M$ whose first row is vector $\bf 1$
(reflecting the fact that the whole set of points of $\EuScript W$ lies in hyperplane
$\langle {\bf y} , {\bf 1}\rangle=0$) and the remaining rows are proportional to vectors
${\boldsymbol \delta}_i$, the rows of the matrix ${\bf WM}^{\sf T}$
all have the form $(0,\pm d_1, \ldots ,\pm d_b)$, with all $d_i >0$,
so that sgn${\bf WM}^{\sf T}=
(0,\pm 1, \ldots, \pm 1)$.  We write
\beq\label{keyeq}
{\bf WM}^{\sf T} = {\bf B}\bf D
\eeq
where $D$ is the diagonal matrix
\beq
{\bf D} =  \mbox{diag}(0,  d_1, \ldots ,  d_b)
\eeq
and the information matrix $\bf B$ has rows of the form $(0, \pm 1, \ldots , \pm 1)$.
From~\eqref{keyeq}, we derive the encoding equation ${\bf W}={\bf BK}$, where the encoding matrix $\bf K$ has the form
\beq
{\bf K} \triangleq {\bf D} {\bf M}^{-\sf T}
\eeq
(Notice that, since all the entries of the first column of $\bf B$ are zero, the first row of $\bf K$ can be replaced by any vector.)
\section{Design examples}\label{examples}
We shall now exhibit a few design examples, while optimization considerations are postponed to next section. One may notice that we have chosen
our operations so that coding and decoding involve only integer numbers. This is not the only choice: for example, one may require all quantities involved in the calculations not to exceed $1$ in absolute value.
\subsection{Example 1 ($b=2$)}\label{example1}
Consider the initial vector ${\bf w}_1=(-1,0,1)$ and the
root permutations  ${\bf w}_2=(-1,1,0)$ and ${\bf w}_3=(1,-1,0)$. From~\eqref{getdelta} we obtain
${\boldsymbol \delta}_1=(0,-1/\sqrt{2},1/\sqrt{2})$ and
${\boldsymbol \delta}_2=(-2/\sqrt{6},1/\sqrt{6},1/\sqrt{6})$, the generator
matrices~\eqref{genmat1}, and hence codebook~\eqref{simpleexample}.
With
\beq
{\bf M} = \left[ \begin{array}{rrr}
1&1&1\\0&-1&1\\2&-1&-1 \end{array}\right]
\eeq
we obtain
\beq
{\bf WM}^{\sf T} = \left[ \begin{array}{rrr}
0&1&-3\\0&-1&-3\\0&-1&3\\0&1&3 \end{array}\right]
\eeq
which yields sgn${\bf WM}^{\sf T}= {\bf B}$, where
\beq
{\bf B} = \left[ \begin{array}{rrr}
0&1&-1\\0&-1&-1\\0&-1&1\\0&1&1 \end{array}\right]
\eeq
With ${\bf D}=\mbox{diag}(0,1,3)$ we also obtain the encoding equation
\beq
{\bf BK} = {\bf BDM}^{-\sf T} = {\bf W}
\eeq
as it should be.
\subsection{Example 2 ($b=3$)}\label{example2}
With $b=3$ and initial vector ${\bf w}_1=(-3,-1,1,3)$, choose the root permutations
${\bf w}_2=(-3,3,1,-1)$, ${\bf w}_3=(-1,-3,3,1)$,  and ${\bf w}_4=(1,-1,-3,3)$. From these we obtain
${\boldsymbol \delta}_1=(0,-1/\sqrt{2},0,1/\sqrt{2})$, ${\boldsymbol \delta}_2=(-1/2,1/2,-1/2,1/2)$, and
${\boldsymbol \delta}_3=(-1/\sqrt{2},0,1/\sqrt{2},0)$, and
hence the following generators of the matrix Coxeter group
isomorphic to ${\mathbb Z}^3_2$:
\beq\label{coxcox}
\begin{aligned}
{\bf O}_1 &= \left[\begin{array}{cccc}
1&0&0&0\\0&0&0&1\\0&0&1&0\\0&1&0&0 \end{array}\right] \\
{\bf O}_2 &= \frac{1}{2} \left[\begin{array}{rrrr}
1&1&-1&1\\1&1&1&-1\\-1&1&1&1 \\ 1&-1&1&1 \\ \end{array}\right] \\
{\bf O}_3 &= \left[\begin{array}{cccc}
0&0&1&0\\0&1&0&0\\1&0&0&0 \\0&0&0&1
\end{array}\right]
\end{aligned}
\eeq
The codebook is
\beq
{\bf W}= {\bf w}_1 \left[\begin{array}{l}
{\bf I} \\ {\bf O}_1 \\ {\bf O}_2 \\ {\bf O}_3 \\
{\bf O}_1{\bf O}_2 \\ {\bf O}_1{\bf O}_3 \\ {\bf O}_2{\bf O}_3 \\{\bf O}_1{\bf O}_2{\bf O}_3
\end{array} \right]
= \left[\begin{array}{rrrr}
-3&	-1&	1&	3\\
-3&	3&	1&	-1\\
-1&	-3&	3&	1\\
1&	-1&	-3&	3\\
-1&	1&	3&	-3\\
1&	3&	-3&	-1\\
3&	-3&	-1&	1\\
3&	1&	-1&	-3
\end{array}\right]
\eeq
The Peterson transformation matrix~\eqref{petmat}
\beq\label{A3D}
{\bf A} = \frac{1}{2}\left[\begin{array}{rrrr}
1&-1&-1&1\\-1&1&-1&1\\-1&-1&1&1\\1&1&1&1
\end{array}\right]
\eeq
yields the $3$-dimensional version of the codebook
\beq
{\bf AW} = \left[\begin{array}{rrrr}
0&2&4&0\\-4&2&0&0\\0&-2&4&0\\4&2&0&0\\
-4&-2&0&0\\0&2&-4&0\\4&-2&0&0\\0&-2&-4&0
\end{array}\right]
\eeq

With
\beq
{\bf M} = \left[\begin{array}{rrrr}
1&	1&	1&	1\\
0& -1&	0&	1\\
-1&	1&	-1&	1\\
-1&	0&	1&	0\\
\end{array}\right]
\eeq
we have that the rows of ${\bf WM}^{\sf T}$ are proportional to $(0,\pm4,\pm 4,\pm 4)$.
Hence, choosing
\beq
{\bf B}= \mbox{sgn}{\bf WM}^{\sf T}
\eeq
and taking ${\bf D}=\mbox{diag}(0,4,4,4)$, we obtain
\beq
{\bf K}= {\bf DM}^{-\sf T} = \left[\begin{array}{rrrr}
0&	0&	0&	0\\
0&	-2&	0&	2\\
-1&	1&	-1&	1\\
-2&	0&	2&	0
\end{array}\right]
\eeq
and ${\bf BK}={\bf W}$, as it should be.
\subsection{Example 3 ($b=3$)}\label{example3}
Consider again $b=3$, and the initial vector ${\bf w}_1=(-3,1,1,1)$. Since the permutations of this vector would yield a codebook with only four points, we assume
that the central inversion matrix $-{\bf I}$ is also an element of the Coxeter group generating the codebook. This is equivalent to assuming that $-{\bf w}_1$ is also a codeword.
Choosing the root permutations $(-1,3,-1,-1)$, $(-1,-1,3,-1)$, and $(-1,-1,-1,3)$,
we obtain the following codebook (called ENRZ in~\cite{primer}):
\beq
{\bf W} = \left[\begin{array}{rrrr}
 -3&1&1&1 \\ -1&3&-1&-1 \\ -1&-1&3&-1 \\ -1&-1&-1&3 \\ 1&1&1&-3 \\ 1&1&-3&1 \\ 1&-3&1&1 \\ 3&-1&-1&-1
\end{array}\right]
\eeq
which can be seen as the union of two PM codebooks, one generated by the four permutations of $(-3,1,1,1)$ and the other generated by the $4$ permutations of $(3,-1,-1,-1)$.\footnote{In a different way, it can be seen as a subset of a Variant-II PM~\cite{slepian2}, which includes not only the permutations of an initial vector, but also the sign changes of its components. This subset should include only the balanced vectors within the Variant-II PM set.}

Using~\eqref{A3D}, the projection $\bf WA$ yields a matrix whose rows are the $8$ vectors of the form
$(\pm 2, \pm 2,\pm 2,0)$, corresponding to a $3$-dimensional cube as shown in Fig.~\ref{cubo}.
\insertfig{h!}{0.5}{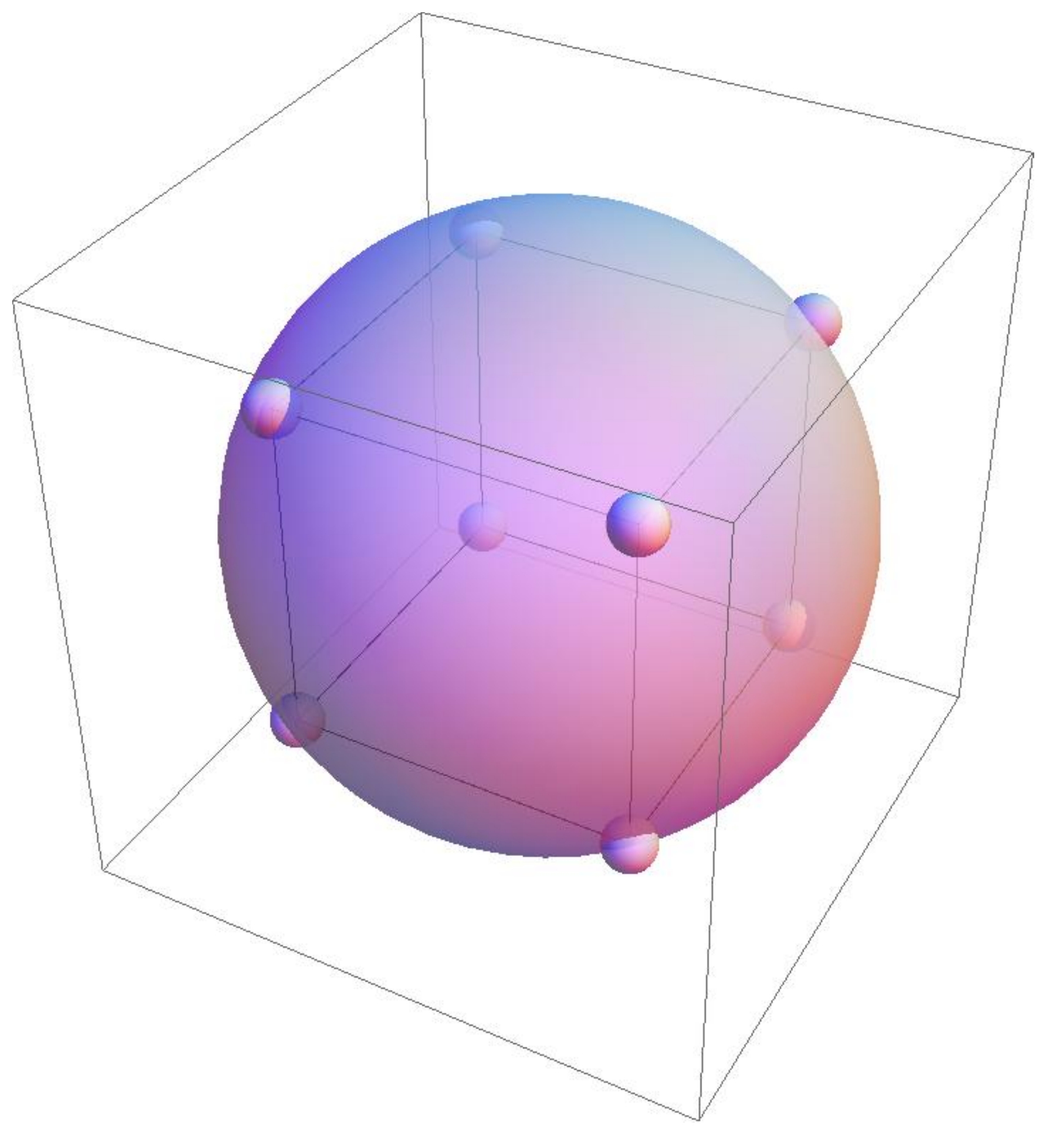}{\sl $3$-dimensional representation of the $8$ points of codebook of Example 3.}{cubo}

$\bf M$ can be given the form of a $4\times 4$ Hadamard matrix:
\beq
{\bf M} = \left[\begin{array}{rrrr}
1&1&1&1 \\ 1&-1&1&-1 \\ 1&1&-1&-1 \\ 1&-1&-1&1
\end{array}\right]
\eeq
which yields a matrix ${\bf WM}^{\sf T}$ whose rows have the form $(0,\pm 4 ,\pm 4, \pm 4)$, and
the encoding matrix may be chosen as ${\bf K}={\bf M}$.
\subsection{Example 4 ($b=3$)}\label{example4}
Choose the initial vector ${\bf w}_1=(-1,0,0,1)$ and the root permutations $(0,-1,0,1)$, $(-1,1,-1,1)$, and $(-1,0,1,0)$. The decoding matrix
\beq
{\bf M} = \left[\begin{array}{rrrrr}
1&1&1&1\\0&-1&0&1\\-1&1&-1&1\\-1&0&1&0
\end{array}\right]
\eeq
yields rows of the matrix ${\bf WM}^{\sf T}$ with the form $(0,\pm 1,\pm 2,\pm 1)$.
The encoding matrix is ${\bf K}=\frac{1}{2} {\bf M}$.
\subsection{Example 5 ($b=4$)}\label{example5}
Take $b=4$, and the initial vector ${\bf w}_1=(-2,-1,0,1,2)$. The root permutations
$(-2,1,0,-1,2)$, $(-1,-2,1,0,2)$, $(-1,0,1,2,-2)$, and $(0,-1,-1,1,2)$ lead to the
decoding matrix
\beq
{\bf M} = \left[\begin{array}{rrrrr}
1&1&1&1&1\\0&-1&0&1&0\\-1&1&-1&1&0\\-1&-1&-1&-1&4\\-1&0&1&0&0
\end{array}\right]
\eeq
which yields rows of the matrix ${\bf WM}^{\sf T}$ with the form $(0,\pm 2,\pm 2,\pm 10, \pm 2)$.
\subsection{Example 6 ($b=5$)}\label{example6}
With $b=5$, choose the initial vector ${\bf w}_1=(1, -1, -3, -1, 1, 3)$ and the
root permutations ${\bf w}_2=(1,1,-3,-1,-1,3)$, ${\bf w}_3=(1,1,-3,-1,3,-1)$, ${\bf w}_4=(-1,-1,1,-3,1,3)$, ${\bf w}_5=(-1, -1,-3,1,1,3)$,
and ${\bf w}_6=(3,-3,-1,1,-1,1)$.
The $32$-word codebook can be decoded using the
matrix
\beq
{\bf M}= \left[\begin{array}{rrrrrr}
1&1&1&1&1&1\\
1&-1&0&0&0&0\\
0&0&0&1&-1&0\\
-1&-1&-1&1&1&1\\
1&1&-2&0&0&0\\
0&0&0&1&1&-2
\end{array}\right]
\eeq
The product ${\bf WM}^{\sf T}$ yields a matrix all of whose rows have the form $(0,\pm 2,\pm 6,\pm 6,\pm 2,\pm 6)$. Encoding is done with ${\bf K}= {\bf M}$.
\section{Consideration of error probabilities}\label{performance}
After addition of white Gaussian noise samples $\sim {\EuScript N}(0,N_0/2)$ independent across wires and transmitted $b$-tuples, the codebook matrix ${\bf W}+{\bf N}$ is received. The detection process is summarized as the calculation of the signs of $({\bf W}+{\bf N}){\bf M}^{\sf T}$. The $j$th symbol of the $i$th source $b$-tuple
is erroneously detected if its polarity is altered by noise, which occurs with probability
\beq
\left( p_{\rm e} \right)_{i,j} = {\mathbb P} \left( n_j < -\left|  \left( {\bf WM}^{\sf T}\right) _{i,j} \right|
\right)
\eeq
where $n_j\sim {\EuScript N}(0,\sigma_j^2)$, and $\sigma_j^2\triangleq (N_0/2)\xi_j^2$  is the $j$th element of the diagonal covariance matrix of the noise term:
\beq\label{powerspectrum}
\begin{aligned}
{\mathbb E}\left[ ({\bf NM}^{\sf T})^{\sf T}({\bf NM}^{\sf T})\right] &= {\bf M}\left[ {\mathbb E}
({\bf N}^{\sf T}{\bf N})\right] {\bf M}^{\sf T} \\
&= \frac{N_0}{2}\mbox{diag}(\xi_1^2, \ldots, \xi_{b+1}^2)
\end{aligned}
\eeq
Thus,
\beq\label{peij}
\left( p_{\rm e} \right)_{i,j} = {\rm Q}\! \left( \frac{ \left| \left( {\bf WM}^{\sf T} \right)_{i,j}\right|}
{\sqrt{N_0/2}\, \xi_j }\right)
\eeq
We define the signal-to-noise ratio $\eta$
observing that the average energy associated with the transmission of a signal $b$-tuple is given by
\beq
{\EuScript E} = \frac{\| {\bf W} \|^2}{2^b}
\eeq
where $\| {\bf W} \|$ denotes the Frobenius norm of matrix $\bf W$.
The energy per bit is consequently
$
{\EuScript E}_b = \EuScript E/b,
$
and the signal-to-noise ratio is
\beq\label{etadef}
\eta \triangleq \frac{\EuScript E_b}{N_0}= \frac{\| {\bf W} \|^2/2^b}{b N_0}
\eeq
Thus, we can rewrite~\eqref{peij} in the form
\beq\label{peijbis}
\left( p_{\rm e} \right)_{i,j} = {\rm Q}\! \left( \alpha_{i,j} \sqrt{2\eta} \right),
\; i=1, \ldots, 2^b, \; j=2,\ldots,b+1
\eeq
where\footnote{Observe that the rows of the matrix $\left|{\bf WM}^{\sf T}\right|$ quantifies the amplitudes of the eye opening before rectification. Notice that having equal columns of matrix $\left| {\bf WM}^{\sf T} \right|$ is not sufficient to have equally protected symbols, as their noise protection also depends on the values $\xi^2_2, \ldots, \xi^2_{(b+1)}$.}
\beq\label{defalphaalt}
\alpha_{i,j}\triangleq  \frac{ \left( \left| {\bf WM}^{\sf T}\right| ({\bf MM}^{\sf T})^{-1/2}\right)_{i,j}}{\sqrt{ \| {\bf W} \|^2/(b2^b)}}
\eeq
Since group codes have the {\em uniform error probability}, i.e.,
the error probability is the same for every transmitted codeword, the values of $\alpha_{i,j}$ do not depend on the value of $i$.
An alternative expression is
\beq\label{altalpha}
\alpha_{1,j}=\sqrt{b} \,\frac{ \| {\bf w}_1-{\bf w}_{j}\|   }{2\cdot \| {\bf w}_1\|}, \qquad j=2,\ldots,b+1,
\eeq
where the denominator is the diameter of the sphere enclosing the codebook vectors.

Using~\eqref{altalpha}, we obtain
\beq\label{sumalpha}
\sum_{j=2}^{b+1} \alpha_{1,j}^2 = b
\eeq
In fact, due to the symmetry of the codebook, if ${\bf w}_1\in {\EuScript W}$ then
also $-{\bf w}_1\in {\EuScript W}$. The distance between ${\bf w}_1$ and $-{\bf w}_1$ is equal to the diameter of the hypersphere on whose surface the codebook points lie, and the vector joining these two points is the longest diagonal of the corresponding orthotope. The squared length of this diagonal equals the sum of the squared lengths of the edges radiating from ${\bf w}_1$, which proves~\eqref{sumalpha}.

Since the Voronoi regions are congruent and bounded by orthogonal hyperplanes, the ML decisions on the individual bits are affected by independent noise samples, and hence
the following exact expression for the average error probability holds:
\beq\label{exactpe}
p_{\rm e} = 1 - \prod_{j=2}^{b+1} \left[ 1 - {\rm Q} \left( \alpha_{1,j} \sqrt{2\eta} \right) \right]
\eeq
We may also observe that~\eqref{exactpe} is minimized, under the constraint~\eqref{sumalpha}, by choosing all the $\alpha_{1,j}$ equal, which corresponds to having the codebook orthotope equal to a hypercube.

From~\eqref{exactpe} we may derive the union upper bound
\beq
p_{\rm e} \leq \sum_{j=2}^{b+1} {\rm Q} \left( \alpha_{1,j} \sqrt{2\eta} \right)
\eeq
and the asymptotic approximation, valid for large signal-to-noise ratios,
\beq
p_{\rm e} \lesssim \nu \, {\rm Q} \left( \alpha_{\rm min} \sqrt{2\eta} \right)
\eeq
where $\alpha_{\rm min}=\min_j   \alpha_{1,j}$, and $\nu$ is the number of $\alpha_{1,j}$
taking value $\alpha_{\rm min}$.

Table~\ref{tablebig} summarizes the values of the $\alpha_{i,j}$ for some line codes.
\begin{table*}
\centering
\caption{\sl Performance of some $(b+1,b)$ Coxeter-group line codes.}\label{tablebig}
\begin{tabular*}{.7\hsize} {@{\extracolsep{\fill}}cccl}
\addlinespace[15pt]
\toprule
$b$ & ${\bf w}_1$ & root permutations &  $\alpha_{1,j}, \;\; j=2,\ldots,b+1$ \\
\toprule
1 & $(1,-1)$ & $(-1,1)$ & $1.$ \\
\toprule
2 & $(-1,0,1)$ & $(-1,1,0)$ & $0.71,1.22$ \\
  & & $(1,-1,0)$ &            \\
\toprule
3 & $(-3,-1,1,3)$ & $(-3,3,1,-1)$ & $0.77,1.1,1.1$ \\
  & & $(-1,-3,3,1)$ & \\
  & & $(1,-1,-3,3)$ & \\
\midrule
3 & $(-1,0,0,1)$ & $(-1,0,1,0)$ & $0.87,0.87,1.22$ \\
  & & $(0,-1,0,1)$  & \\
  & & $(0,1,-1,0)$  & \\
\midrule
3 & $(-3,1,1,1)$ & $(-1,3,-1,-1)$ & $1,1,1$ \\
  & & $(-1,-1,3,-1)$ & \\
  & & $(-1,-1,-1,3)$ & \\
\toprule
4 & $(-2,-1,0,1,2)$ & $(-2,1,0,-1,2)$ & $0.63,0.89,0.89,1.41$ \\
  & & $(-1,-2,1,0,2)$ & \\
  & & $(-1,0,1,2,-2)$ & \\
  & & $(0,-1,-2,1,2)$ & \\
\toprule
5 & $(1,-1,3,-3,5,-5)$ & $(-1,1,5,-5,3,-3)$ & $0.66,0.76,0.76,1.31,1.31$ \\
  & & $(3,-3,1,-5,5,-1)$ & \\
  & & $(3,-3,5,-1,1,-5)$ & \\
  & & $(3,5,-3,-1,1,-5)$ & \\
  & & $(-5,-3,1,3,5,-1)$ & \\
\midrule
5 & $(-2,-1,0,0,1,2)$ & $(-2,-1,2,0,1,0)$ & $0.71,1.0,1.0,1.0,1.22$ \\
  & & $(-2,0,-1,0,2,1)$ & \\
  & & $(-2,1,0,0,-1,2)$ & \\
  & & $(0,-2,-1,2,0,1)$ &\\
  & & $(0,-1,0,-2,1,2)$ & \\
\midrule
5 & $(1,-1,-3,-1,1,3)$ & $(1,1,-3,-1,-1,3)$  & $0.67,0.67,1.17,1.17,1.17$\\
  & & $(1,1,-3,-1,3,-1)$ & \\
  & & $(-1,-1,1,-3,1,3)$ & \\
  & & $(-1,-1,-3,1,1,3)$ & \\
  & & $(3,-3,-1,1,-1,1)$ & \\
\toprule
\end{tabular*}
\end{table*}
%

%
\section{Optimization of the codebook}\label{optimization}
A natural and common optimization criterion, based on the performance at large values of signal-to-noise ratios, is the maximization of the minimum Euclidean distance of the codebook, i.e., of
\beq\label{dmin}
d_{\rm min} = \min _{\bf O} \| {\bf w}_1 - {\bf O}{\bf w}_1 \|
\eeq
where $\bf O$ runs through the matrices representing the Coxeter group chosen for the codebook design.
The choice between two line codes with the same $d_{\rm min}$ may be based on the
second smallest Euclidean distance, etc.
Since the design criterion described in Section~\ref{theory} generates a codebook which is a subset of a PM set,
the minimum distance of the latter turns out to be a lower bound on~\eqref{dmin}.
\subsection{Choosing the initial vector}
The first constraint on the choice of ${\bf w}_1$ comes from the observation that, due to
the linearity of the encoder, if ${\bf w}\in {\EuScript W}$ then also $-{\bf w}\in {\EuScript W}$. A sufficient condition for this to occur is to force $-{\bf w}$ to be a permutation of ${\bf w}$, which is obtained from an initial vector such that
its nonzero components occur in pairs including positive and negative values. All the examples in Table~\ref{tablebig} satisfy this condition, with the only exception of the entry described in Example~\ref{example3}.

Further, it seems reasonable to start from an original PM set having the
largest possible minimum distance.\footnote{One should keep in mind that it may occur that the minimum distance of the line code be larger than that of the original PM set.} Using the notations of~\cite{slepian}, the initial vector for the generation of a PM set has components $\mu_1, \ldots, \mu_k$, each being different and repeated
$m_1, \ldots, m_k$ times, respectively. It was proved in~\cite{bigeli} that for optimality the $\mu_i$ must be equally spaced (i.e., $\mu_{i+1}-\mu_i$ is a constant). Moreover, if $m_1, \ldots, m_k$ are given,
then the optimum combination of $\mu$s and $m$s consists of pairing the smallest $m$ with the smallest $\mu$, the second smallest $m$ with the largest $\mu$, the third smallest $m$ with the second smallest $\mu$, and so forth. Thus, the optimization of a PM set is complete once the $m$s are chosen in an optimum way. In~\cite{ingemarsson1,ingemarsson2}, Ingemarsson has advocated a choice of the $m$s which makes the amplitudes of the initial vector have a sampled Gaussian distribution (an idea that was used in~\cite{slepian}). However, the solution of~\cite{ingemarsson1,ingemarsson2} may not be optimum, as the search was restricted to initial vectors satisfying a certain symmetry~\cite{mittelholzer,mitlah}.
A numerical optimization algorithm was derived by Karlof~\cite{karlof1,karlof2}, while tables of optimum PM sets in low dimensions are exhibited in~\cite{ericson}. For small values of $b$, a sensible choice consists of checking all the partitions $(m_1, m_2, \ldots , m_r)$ of the number of components
of ${\bf w}_1$, as mentioned in Section~\ref{motivation}, and choosing the partition yielding the best code.

In our design, for $d_{\rm min}$ optimization, we examine all the partitions of $(b+1)$ in the form
$b+1=m_1+\ldots+m_k$, and derive for each of them a codebook under the assumption
of equally spaced $\mu_i$. This is shown in Examples~\ref{example2} to~\ref{example4}, where the partitions
$4=1+1+1+1$, $4=1+2+1$, and $4=3+1$ were considered. Notice also that some of the partitions
may not lead to a codebook satisfying our constraints: for example, the partition $4=2+2$ generates a PM set with $4!/(2!2!)=6$ vectors, which cannot be used to generate a codebook with $2^b=8$ vectors as needed.
\subsection{Choosing the root permutations}
Once ${\bf w}_1$ has been chosen, the Coxeter matrix group has to be generated, which is obtained, as described in Section~\ref{theory},
by taking $b$ additional permutations ${\bf w}_i$, $i=2, \ldots, b+1$, (the root permutations) such that the $b$ difference vectors $({\bf w}_1- {\bf w}_i)$ are mutually orthogonal. These vectors correspond to the $b$ edges of a $b$-dimensional orthotope having ${\bf w}_1$ as a vertex. We observe first that
in some cases such permutations may not exist. For example, the initial vector
${\bf w}_1=(-1,0,0,0,1)$ originates a PM set with $20$ vectors, from which $4$ orthogonal difference vectors cannot be found. In other cases, more than one choice of root permutations is available, as shown graphically in Fig.~\ref{figura_esagono_con_cricche} for the simple case $b=2$. From this figure it is seen that the two choices are equivalent, as they give rise to congruent orthotopes, but this may not be the case for $b>2$.
%
\insertfig{h!}{0.4}{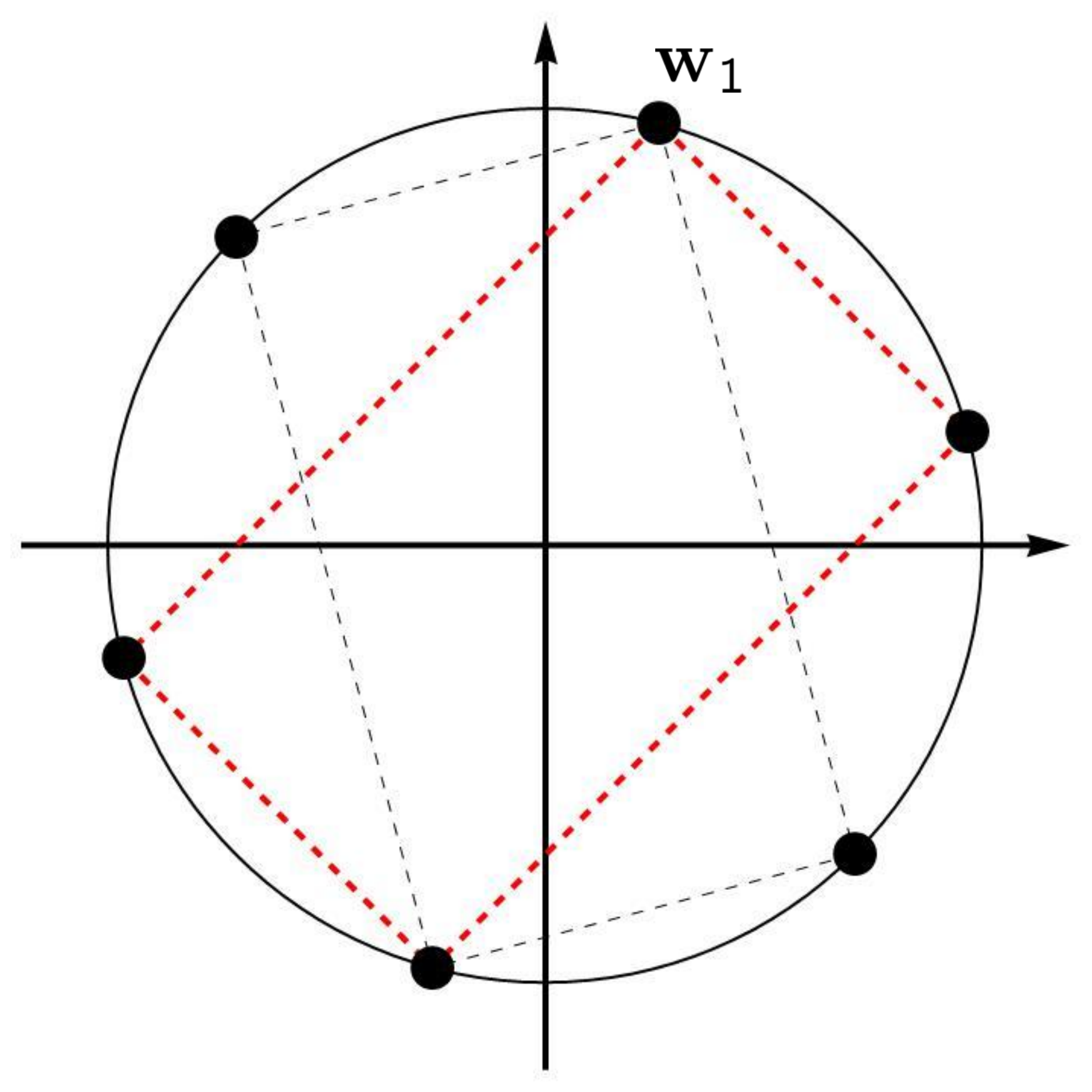}{\sl Two possible choices of root permutations leading to equivalent codebooks with $b=2$.}{figura_esagono_con_cricche}

In other cases the different choices of root permutations yield codebooks with different performance, as revealed by the Euclidean distances from ${\bf w}_1$ to the other root permutations. A simple algorithm listing all the choices of root permutations and their quality consists of the following.
First, form the matrix ${\bm \Delta}$ whose rows are the differences between
${\bf w}_1$ and all its permutations. The Gram matrix ${\bm \Delta}{\bm \Delta}^{\sf T}$ has zero in all
entries corresponding to a pair of orthogonal differences. From this matrix we can generate the
incidence matrix of a graph whose vertices are those differences, and the edges join
vertex pairs corresponding to orthogonal differences. A clique of this graph is a subset of the vertices corresponding to mutually orthogonal differences. The maximum number of vertices in such a clique is $b$, and the clique is called maximal. Thus, the problem of choosing a set of root permutations is tantamount to that of choosing a maximal clique with the largest minimum norm of the orthogonal differences in it (as mentioned before, if two cliques have the same minimum norm, we choose the one whose second smallest distance is the largest, etc.).

For example, the initial vector ${\bf w}_1=(1,-1,-3,-1,1,3)$ of Example~\ref{example6} has $180$ permutations and $24$ maximal cliques. The best clique under our criterion yields the values of $\alpha_{1,j}$ listed in the last entry of Table~\ref{tablebig}. A related line code, using the same ${\bf w}_1$ and exhibited in~\cite[Table 2]{patshok1}, yields a slightly inferior performance (the values of $\alpha_{1,j}$ are $0.67$, $0.67$, $0.67$, $0.95$, and $1.65$).
\subsection{Removing the PM and integer-number constraint}
The designs done in the previous sections were based on the constraint of a codebook being a subset
of a PM set, as this choice reduces the cardinality of the set of the signal amplitudes in each wire. This is a convenient choice, because a limited number of amplitudes implies
a limited number of current or voltage sources needed to implement the encoder. In addition, one can deal only with integer amplitudes, thus increasing the accuracy of the implementation as rounding becomes unnecessary. The downside of this choice is that the vertices of the codebook orthotope are constrained to a subset of those of the polytope of the original PM set (a {\em semiregular} polytope, see~\cite{slepian}). If this constraint is removed, after a Coxeter matrix group is generated, one may choose the optimum initial vector as indicated in~\cite{mitlah}, that is, being at the same distance from every plane bounding the fundamental region of the Coxeter group in which ${\bf w}_1$ lies.~\footnote{We recall that the fundamental region of a matrix group is a connected region of the space such that no point in its interior can be obtained as ${\bf Ow}_1$, where $\bf O$ is any matrix of the group. For a precise definition see, e.g.,~\cite{mitlah}.}
\insertfig{h!}{0.4}{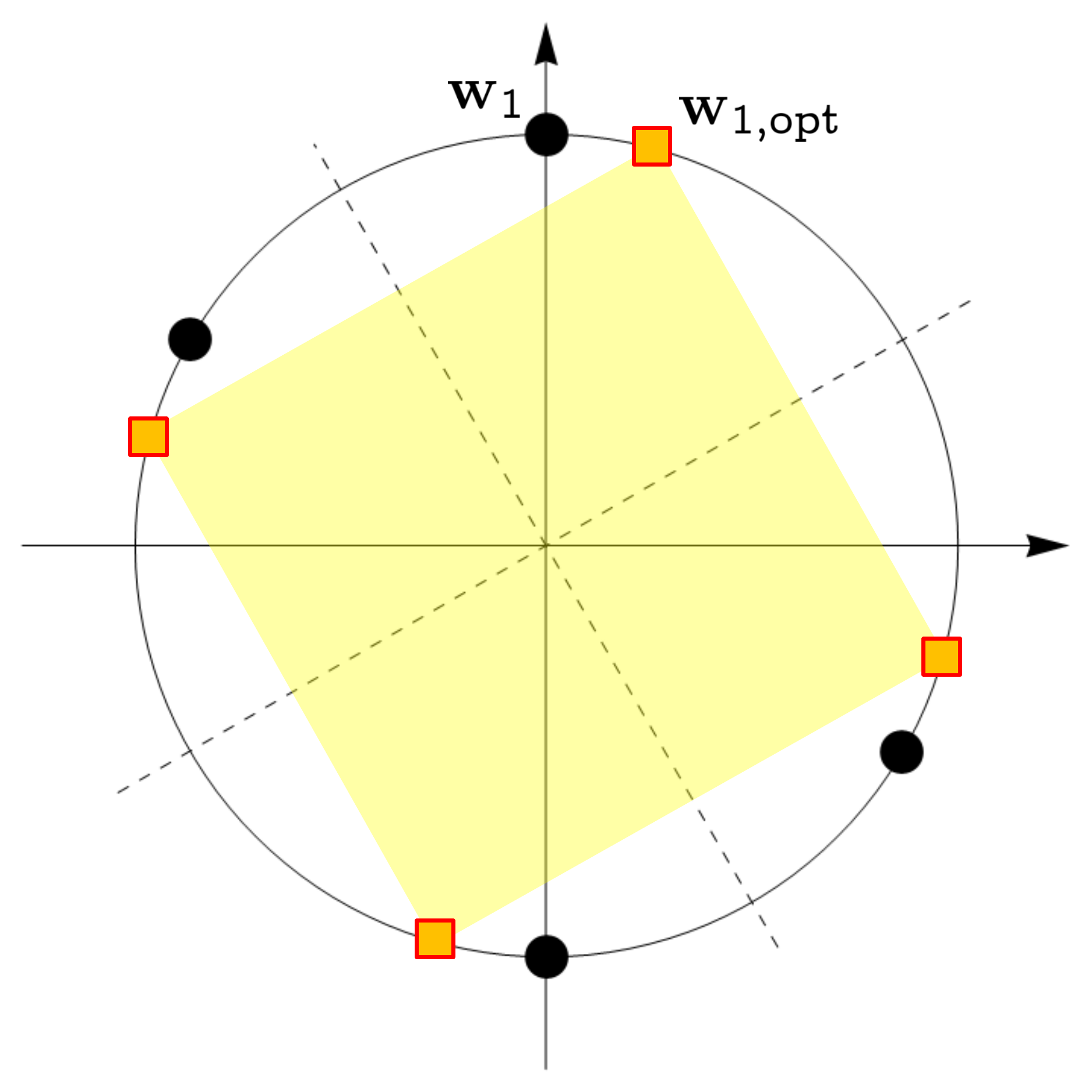}{\sl Comparison of the line codes with $b=2$ obtained
from the same matrix Coxeter group, as applied to ${\bf w}_1$ (see the entry with $b=2$ in Table~\ref{tablebig}) and to ${\bf w}_{1, \rm opt}$.}{figuronabisottimo}
In general, we have
\beq
{\bf w}_{1, \rm opt} = \sum \frac{{\bf w}_1-{\bf w}_i }{\| {\bf w}_1-{\bf w}_i \|}
\eeq
where the sum runs through the set of root permutations. The resulting codebook has the shape of a hypercube, which yields $\alpha_{1,j}=1$ for $j=2, \ldots, b+1$. Fig.~\ref{figuronabisottimo} shows a geometric representation of the codebook obtained with the choice of the optimum initial vector.

For suboptimum design, one may use an integer-value approximation of the optimum initial vector, which could lead to a codebook shape close to a hypercube,
possibly at the price of a larger number of values of the codeword components.
For illustration, consider $b=3$ and the first set of root permutations of Table~\ref{tablebig}. The matrix representation of the Coxeter group in this case yields the optimum initial vector
${\bf w}_{1,\rm opt}= (-1/2-\sqrt{2}/2,1/2-\sqrt{2}/2,-1/2+\sqrt{2}/2,1/2+\sqrt{2}/2)$, and a codebook $\bf W$ whose rows are permutations of ${\bf w}_{1,\rm opt}$. Using the suboptimum initial vector $(-6,-1,1,6)\approx 5\times{\bf w}_{1,\rm opt}$, one obtains
$\alpha_{1,j}\approx 1$.
\section{Miscellaneous remarks}\label{miscel}
\begin{dingautolist}{192}
\item
In~\cite{mitlah,petnatfos}, group codes generated by Coxeter groups were studied. The constraint
of having Voronoi regions bounded by orthogonal hyperplanes, and hence allowing an exceedingly simple simple ML detection, was not considered. The designs in~\cite{mitlah} were optimized by choosing an initial vector in the center of a fundamental region of the Coxeter group.
\item
A topic related to the codebooks examined in this paper is the study of constant-weight codes. These satisfy the equal-energy condition, while their words may not be balanced in the sense of this paper. See, e.g.,~\cite{pelelmtalbos,talbos} and the references therein.
\item
Introduction of error-control capabilities can be obtained by suitably decreasing the wire efficiency
and using standard linear codes, as advocated in~\cite{patcronie2}.
\item
It should be noticed that the subset of permutations leading to a line code constructed using a Coxeter group does not necessarily form a subgroup of matrices of the natural representation of the symmetric group. In fact, although the generating matrices of a representation of the Coxeter group, as applied to ${\bf w}_1$, yield permutations of that vector, these may not be all permutation matrices.
As an example, the ``square'' code of Fig.~\ref{figuronabisottimo}, generated by the optimum ${\bf w}_1$ approximately equal to $(-0.8, -0.3, 1.1)$, yields the codebook
\beq
{\bf W} = \left[ \begin{array}{rrr}
-0.8 &	-0.3 &	1.1 \\
-0.8 &	1.1 &	-0.3 \\
0.8 & -1.1 & 0.3 \\
0.8 & 0.3 & -1.1 \\
\end{array}\right]
\eeq
As another example, with the design of last entry in Table~\ref{tablebig} the optimum initial vector can be found to be $(1,-1,-\sqrt{3},-1,1,\sqrt{3})$ yields a codebook whose words have again an increased alphabet size and are not permutations of the initial vector.
\end{dingautolist}
\section{Conclusions}
Expanding on the work described in~\cite{abbasfar,primer,patcronie2,patshok1}, we have developed an algebraic method
for generating line codes for parallel transmission that have many of the properties of differential signaling. These are group codes generated by a matrix representation of a Coxeter group. Performance evaluation is also discussed, and a number of design examples are exhibited (some of which are new, or improve upon known codes), along with some consideration of optimum codes.
\section*{Acknowledgments}
The authors gratefully acknowledge the comments of Michele Elia on an early version of this paper.
The work of Ezio Biglieri  is supported by Project TEC2015-66228-P, while that of
Emanuele Viterbo is supported by ARC under Grant Discovery Project No. DP160101077.
%
%


\begin{thebibliography}{99}
%
\bibitem{abbasfar}
A.\ Abbasfar, ``Generalized differential vector signaling,''
{\em IEEE International Conference on Communications (ICC 2009)},
Dresden, Germany,  June 14--18, 2009.
%

\bibitem{bigeli}
E.\ Biglieri and M.\ Elia, ``Optimum permutation modulation codes
and their asymptotic performance,''
{\em IEEE Trans.\ Inform.\ Theory}, vol.\ 22, pp.\ 751--753, November 1976.
%

\bibitem{bigvit1}
E.\ Biglieri and E.\ Viterbo, ``Line coding for differential vector signaling,''
{\em ITA Workshop}, San Diego, CA, February 12--17, 2017.
%

\bibitem{bigvit2}
E.\ Biglieri and E.\ Viterbo, ``Geometrically uniform differential vector signaling schemes,''
{\em IEEE ISIT 2017}, Aachen, Germany, June 25--30, 2017. 
%


\bibitem{bjobre}
A.\ Bjorner and F.\ Brenti, {\em Combinatorics of Reflection Groups}. New York: Springer, 2005.
%

\bibitem{coxeter}
H.\ S.\ M.\ Coxeter, ``Discrete groups generated by reflections,''
{\em Annals of Mathematics}, vol.\ 35, no.\ 3, pp.\ 588--621, July 1934.
%

\bibitem{coxeterbook}
H.\ S.\ M.\ Coxeter, {\em Regular Polytopes}. London, U.K.: Methuen, 1948.
%

%

\bibitem{ericson}
T.\ Ericson, ``Permutation codes,'' {\em Rapport de Recherche INRIA} n.\ 2109, November 1993.
%

%

\bibitem{primer}
``A primer on chord signaling,'' Downloaded August 2016 from website \url{https://www.kandou.com/technology/coding}.
%

\bibitem{patcronie1}
H.\ Cronie and A.\ Shokrollahi, ``Orthogonal differential vector signaling,''
Patent Application US 2011\slash 0268225~A1, November 3, 2011.
%

\bibitem{patcronie2}
H.\ Cronie and A.\ Shokrollahi, ``Power and pin efficient chip-to-chip communications with common-mode rejection and SSO resilience,'' Patent Application US 2011\slash 0302478~A1, December 8, 2011.
%

\bibitem{groben}
L.\ C.\ Grove and C.\ T.\ Benson, {\em Finite Reflection Groups}. New York: Springer, 1985.
%

\bibitem{humphreys}
J.\ E.\ Humphreys, {\em Reflection Groups and Coxeter Groups}. Cambridge, U.K.: Cambridge Univ.\ Press, 1985.
%

\bibitem{ingemarsson1}
I.\ Ingemarsson, ``Optimized permutation modulation,''
{\em IEEE Trans.\ Inform.\ Theory}, vol.\ 36, no.\ 5, pp.\ 1098--1100, September 1990.
%

\bibitem{ingemarsson2}
I.\ Ingemarsson, ``Group codes for the Gaussian channel,'' in: G.\ Einarsson {\em et al.},
{\em Topics in Coding Theory}. Lecture Notes in Control and Information Sciences,vol.\ 128. New York: Springer Verlag, 1989, pp.\ 73--108.
%

\bibitem{karlof1}
J.\ Karlof, ``Permutation codes for the Gaussian channel,''
{\em IEEE Trans.\ Inform.\ Theory}, vol.\ 35, no.\ 4, pp.\ 726--732, July 1989.
%

\bibitem{karlof2}
J.\ Karlof and Y.\ O.\ Chang, ``Optimal permutation codes for the Gaussian channle,''
vol.\ 43, no.\ 1, pp.\ 356--358, January 1997.
%

\bibitem{mittelholzer}
T.\ Mittelholzer, ``Construction and decdoing of optimal group codes from finite reflection groups,''
in: {\em Communications and Cryptography: Two Sides of One Tapestry,} R.\ E.\ Blahut {\em t al.}, Eds., Norwell, MA: Kluwer, 1994.
%

\bibitem{mitlah}
T.\ Mittelholzer and J.\ Lahtonen, ``Group codes generated by finite reflection groups,''
{\em IEEE Trans.\  Inform.\ Theory}, vol.\ 42, no.\ 2, pp.\ 519--528, March 1996.
%

\bibitem{ohetal}
D. Oh, F.\ Ware, W.\ Kim, J.-H.\ Kim, J.\ Wilson, L.\ Luo, J.\ Kizer, R.\ Schmitt, C.\ Yuan, and J.\ Eble, ``Pseudo-differential signaling scheme based on 4b\slash 6b multiwire code,''
{\em IEEE 17th Topical Meeting on Electrical Performance of Electronic Packaging (EPEP'08)},
San Jos\'e, CA, pp.\ 29--32, October 27--29, 2008.
%

\bibitem{pelelmtalbos}
D.\ Pelusi, S.\ Elmougy, L.\ G.\ Tallini, and B.\ Bose,
``$m$-ary balanced codes with parallel decoding,''
{\em IEEE Trans.\ Inform.\ Theory}, vol.\ 61, n.\ 6, pp.\ 3251--3264, June 2015.
%

\bibitem{petnatfos}
W.\ W.\ Peterson, J.\ B.\ Nation, and M.\ P.\ Fossorier, ``Reflection group codes and their
decoding,'' vol.\ 56, no.\ 12, pp.\ 6273--6293, December 2010.
%

\bibitem{peterson}
W.\ W.\ Peterson, ``A note on permutation modulation,''
{\em IEEE Trans.\ Inform.\ theory}, vol.\ 43, no.\ 1, pp.\ 359-360, January 1997.
%

%

%

%


\bibitem{patshok1}
A.\ Shokrollahi and R.\ Ulrich, ``Vector signaling codes with increased signal to noise characteristics,''
Patent Application US 2016\slash 0013954~A1, January 14, 2016.
%

\bibitem{slepian}
D.\ Slepian, ``Permutation modulation,'' {\em IEEE Proceedings},
pp.\ 228--236, March 1965.
%

\bibitem{slepian2}
D.\ Slepian, ``Group codes for the Gaussian channel,''
{\em Bell System Technical Journal}, vol.\ 47, pp.\ 575--602, 1968.
%

\bibitem{talbos}
L.\ Tallini and B.\ Bose, ``Transmission time analysis for parallel asynchronous communication scheme,'' {\em IEEE Trans.\ Computers}, vol.\ 32, no.\ 5, pp.\ 558--571.
%


\bibitem{vitperm}
E.\ Viterbo, ``Permutation Codes,'' in {\em Encyclopedia of Telecommunications}.
J.\ Proakis (ed.), John Wiley \& Sons, 2002.
%

\end{thebibliography}
\end{document}